\documentstyle[11pt,newpasp,twoside]{article}
\index{summary}
\index{instructions}
\index{template}

\def\edcomment#1{\iffalse\marginpar{\raggedright\sl#1\/}\else\relax\fi}
\reversemarginpar

\begin{document}
\title{Pulsar Wind Nebulae as High Energy $\gamma-$Ray Emitters}
 \author{Mallory S. E. Roberts \altaffilmark{1}}
\affil{McGill University, 3600 University St.,  Montreal, QC, Canada}
\author{Bryan M. Gaensler}
\affil{Harvard-Smithsonian Center for Astrophysics, Cambridge, MA, USA}
\author{Roger W. Romani}
\affil{Stanford University, Stanford, CA, USA }
\altaffiltext{1}{Massachusetts Institute of Technology, Cambridge, MA, USA}

\begin{abstract}
There is growing evidence that some variable high-energy $\gamma-$ray 
sources are associated with pulsar wind
nebulae. We review the current status of X-ray and radio studies of the sources most likely to be significant emitters above
100 MeV. The assumption of the $\gamma-$radiation arising from synchrotron
processes puts a lower limit on the magnetic field and suggests there may 
be significant doppler boosting causing the asymmetries seen in the 
morphology of these nebulae.  However, these sources are also in close
proximity to molecular clouds, suggesting that variable 
$\gamma-$ray emission may arise from a ram pressure
confined pulsar wind nebula in a dense 
medium.
\end{abstract}

\section{Introduction}

For over 30 years, the nature of the majority of high-energy
($E > 100$ MeV) $\gamma-$ray sources in the Galactic plane has
been a mystery. The only firmly identified source class has been
spin-powered pulsars, whose phase-averaged emission appears constant. However,
the $EGRET$ data clearly show that there is at least one
class of variable emitters in the Galaxy (eg. Tompkins 1999).
Several suggestions as to the nature of these objects have been put forward
but no strong observational evidence has previously supported a 
particular source class.

However, the variability timescale of a 
few months to a few years for sources with emission above 100 MeV
gives us some clues to their nature. First, a maximum 
emission region size can
be associated with a light crossing time of a few months or less.
For typical Galactic distances of a few kiloparsecs,
this implies angular extents on the sky 
$\la 1^{\prime}$. An acceleration mechanism that 
can generate very high energy particles in a short period must also be 
present. 
One known source class fitting these criteria are pulsar wind nebulae
(PWN).

\section{The Candidate PWN}

There are around 10 X-ray and radio PWN discovered so far around known 
pulsars (eg. Gaensler et al. 2000; Chevalier 2000), with a
similar number of good candidates without a detected pulsar.
Remarkably, around half of these sources are associated
with $\gamma-$ray pulsars or coincident with unidentified $EGRET$ sources. 
Only the oldest $\gamma-$ray pulsars, such as
Geminga and PSR B1055-52, show no sign of having PWN.
Therefore, it might be expected that other $\gamma-$ray pulsars
might reveal themselves through PWN.

None of the known $\gamma-$ray pulsars are significantly 
variable as measured above 100 MeV using the $\tau$ statistic 
(Tompkins 1999).
However, in the 70-150 MeV range, the unpulsed flux of the Crab was observed
to vary (de Jager et al. 1996), probably a synchrotron
emission component from the inner nebula, 
perhaps associated with the variable
optical and X-ray wisps (Scargle 1967, and Hester etc. these proceedings),
with a variable exponential cutoff at around 25 MeV. 
Chandra has also revealed spatial variability of the X-ray emission 
on a timescale of a few months in the Vela and G11.2$-$0.3 PWN 
(Kargaltsev et al., Roberts et al. these proceedings). Therefore, 
high-energy variability on the appropriate time and spatial scales appears
to be a common feature of PWN, and the question becomes that
of what energy the $\gamma-$ray spectral cut-off is and why. 

In a recent flux-limited survey of the brightest sources of emission above
1 GeV, Roberts, Romani, and Kawai (2001) noted that the four unidentified
sources showing the most variability above 100 MeV 
(out of 20 galactic sources in their sample measured for variability by
Tompkins 1999) contained the brightest new sources of 
extended hard X-ray emission found in the survey and the previously
known PWN around PSR B1853+01 in the supernova remnant W44. The supernova
remnant W44 is one of a class of remnants which is centrally filled with 
thermal X-rays, implying a relatively dense interior.
The radio PWN around PSR B1853+01 is a plume reaching only to the north of
the pulsar, suggesting a ram-pressure confined wind
(Slane et al. 1997). The hard X-ray emission 
is also offset to the north.

A second variable source, GeV J1417$-$6100, coincides with
two extended hard X-ray sources. Radio studies of these sources show they
are within a complex of radio emission called the Kookaburra.
One is coincident with the Rabbit radio nebula, whose morphology, 
relatively flat (spectral index $\alpha=-0.25$) 
spectrum, and high polarization all mark it as a probable PWN 
(Roberts et al. 1999). The polarized emission has a trail-like 
morphology, the head of which is coincident
with an X-ray peak seen in the $ASCA$ SIS.
The second X-ray source is at the southern edge of
the upper wing of the Kookaburra, whose high ratio of infrared to radio
emission, excess polarization,  and non-thermal spectrum mark it as a 
likely supernova remnant, although whether it is partially or
fully a result of a pulsar wind is difficult to determine with the 
existing data (Roberts, Romani \& Johnston 2001). 
This second X-ray source has recently been found to
contain the high $\dot E$ pulsar PSR J1420$-$6048 (D'Amico et al. 2001), 
and is therefore an X-ray and probably radio PWN. 

A third variable source, GeV J1825$-$1310, contains a newly discovered
extended hard X-ray source with no obvious bright radio counterpart.
We have imaged the region with the VLA at 20~cm, and find 
much excess radio emission in the region, although precise measurements
are difficult due to several nearby bright HII regions. 
The MSX satellite 8.3~$\mu$m image shows the infrared emission
in the region to be partially correlated with the radio emission, but
not clearly so in the vicinity of the X-ray nebula.  
Although the X-ray nebula in GeV J1825$-$1310 does not show all of the 
observational signatures of a PWN, other interpretations of its
nature seem even less well justified.   
  
The fourth variable source, GeV J1809$-$2327, contains an extended 
X-ray complex partially coincident with stars in an OB association.
It is contained within the Lynds 227 dark nebula, and appears to be
interacting with a cloud imaged in CO (Oka et al. 1999). A recent short
Chandra ACIS exposure resolves much of the emission into point
sources coincident with stars in the OB association and embedded in a
thermal radio cloud, but it also reveals
a point source embedded in diffuse emission somewhat to the North of the 
stellar cluster with no optical counterpart but with a short bright
trail of emission leading from it (Braje et al. 2001). This latter source
is at the tip of a funnel-shaped radio nebula whose high polarization and
spectral index strongly suggest it is a PWN (Roberts, Gaensler \& 
Romani 2002). 
With this GeV source also coinciding with a likely X-ray
and radio PWN, we are led to the
hypothesis that some of the variable $EGRET$ sources are PWN.

\section{Emission Mechanisms}

de Jager et al. (1996) argue that a general upper limit
to the cut-off energy of synchrotron emission from shock accelerated
electrons can be derived by balancing the acceleration timescale (equal
to the electron gyroperiod across a strong shock) with the synchrotron 
loss timescale. The resulting cutoff energy is:

$$E_{max}=25(D\alpha/{\rm sin}\theta)\,{\rm MeV}$$

\noindent
Here $D$ is the Doppler boost factor, $\theta$ is the electron pitch
angle, and $\alpha < 1$. If this is the true limit on emission from shock generated
electrons, then to reach energies of several hundred MeV, either the
Doppler factor is large or the pitch angle is small. If the one-sided 
nature of some of the nebulae is due to large Doppler boost factors 
then perhaps the emission can reach the required energies.

On the other hand, 
Gallant \& Arons (1994) suggest that ions play an important
role in the generation of wisps, and the spatial and time
scales are related to the ion cyclotron radius, with emission coming
from electrons coupled to the ions.  For ion-pumped shock acceleration, 
the maximum emission energy is proportional to the ambient magnetic field
$E_{max}\propto B\gamma^2 \propto B \dot E$.
%\begin{figure}[h]
%\psfig{file=.eps,width=3cm}
%\psfig{file=.eps,width=3cm}
%\caption{}
%\end{figure}
%
%\begin{figure}[h]
%\psfig{file=.eps,width=3cm}
%\psfig{file=.eps,width=3cm}
%\caption{}
%\end{figure}
Assuming that the variable $\gamma-$ray emission is synchrotron
radiation, we can estimate a lower limit on the
magnetic field in the emitting region. It is reasonable to
assume that the synchrotron cooling timescale must be  shorter than the
variability time scale ($t_{cool} \le t_{var}$).
Therefore, a lower limit on the magnetic field is given by:

$$B\ge 10^{-4} \left( {10{\rm MeV}\over E}\right)^{1/3} t_{var}^{-2/3}$$

\noindent
For $E=100$ MeV and $t_{var}= 0.2-2$ yr., the magnetic
field is $B > 0.3 - 1.4 \times 10^{-4}$ G, similar to the
inferred magnetic fields in the Crab and Vela PWN.

Observationally, the $\gamma-$ray emitting candidates seem to
be in dense environments (near or in molecular clouds or overdense SNR),
and may all be bow-shock nebulae. The magnetic field could
then be compressed and amplified if the pulsars have high velocities:

$$ E_{max} \propto B\dot E \propto \sqrt{\rho}v\dot E $$

\noindent
although to increase the cutoff to many times that of the Crab from
pulsars whose spin-down energy is one or two orders of magnitude 
smaller may be quite difficult.

Alternatively, perhaps other emission mechanisms are important. 
Bremmstrahlung radiation from the pulsar wind striking an
ambient molecular cloud (Oka et al. 1999) has been suggested in the case of
GeV J1809$-$2327, although it may be too inefficient if we require the 
emission region to be as small as implied by the variability time scale. 
Inverse Compton scattering is also an attractive alternative. However, the
cooling timescale of the electrons involved for IC emission in the $EGRET$ 
waveband is much longer than the variability timescale, implying the
density of low-energy photons that are upscattered at the emission site 
would need to be modulated in some manner.

Ultimately, $GLAST$ and $AGILE$ will make unambiguous source identifications,
and will be able to verify the variability seen by $EGRET$, allowing
correlations to be made with the low-energy spatial variability. However, 
if a significant fraction of the emission is truly variable, the pulse
fraction of these sources is small, and the detection of any pulsed 
$\gamma-$ray emission through blind searches is unlikely to succeed. 
Therefore, X-ray and radio pulse detections will still be necessary
to detect $\gamma-$ray pulsations and determine the energetics
of these systems.

\end{document}